\documentclass[a4paper]{jpconf}
\usepackage{graphicx}
\begin{document}
\title{Physics problems and instructional strategies for developing social networks in university classrooms}

\author{Javier Pulgar$^{1,2}$, Carlos Rios$^3$ and Cristian Candia$^{4,5}$}

\address{$^1$ Departamento de F\'isica, Universidad del B\'io B\'io, Concepci\'on, Avda. Collao 1202, Chile}
\address{$^2$ Gevirtz School of Education, University of California Santa Barbara, CA, 93106-9490, USA}
\address{$^3$ Departamento de Ense\~nanza de las Ciencias B\'asicas, Universidad Cat\'olica del Norte, Larrondo 1281, Coquimbo, Chile}
\address{$^4$ Centro de Investigaci\'on de Complejidad Social, Facultad de Gobierno, Universidad del Desarrollo, Avda. Las Condes 12461, Santiago, Chile}
\address{$^5$ Collective Learning Group, MIT Media Lab, Massachusetts Institute of Technology, MA, 02139-42, USA
}
\ead{jpulgar@ubiobio.cl, carlos.rios@ucn.cl, ccandiav@mit.edu}

\begin{abstract}
In this study we explored the extent to which problems and instructional strategies affect social cohesion and interactions for information seeking in physics classrooms. Three sections of a mechanics physics course taught at a Chilean University in Coquimbo were investigated. Each section had a weekly problem-solving session using different sets of well and/or ill-structured problems (i.e., algebra-based and open-ended problems respectively), as well as instructional strategies for guiding the problem solving sessions. Data was collected on networks of information seeking and perceptions of good physics students, during a problem solving session. We used social network analysis (SNA) for constructing variables while conducting the study. Results suggest that the teaching and learning strategies to guide problem solving of well and ill-structured problems yield different levels of social interaction among classmates, and significant levels of activity in seeking out information for learning and problem-solving. While strategies for guiding problem solving lend to significant differences for network connectivity, well and ill-structured physics problems predict similar levels of social activity. 
\end{abstract}

\section{Introduction}
Student networks play a key role when promoting not only learning opportunities, but a sense of community and identity, all of which are necessary to achieve success at a university \cite{1_Zwolak,2_Stadtfeld}. In this study, we explored the learning conditions that foster (or hinder) students social systems, by paying close attention to classroom networks and understanding their importance for academic success and social integration. This work is part of a larger research project that attempts to find appropriate physics teaching strategies and activities for social competencies, team work, and deeper learning. We investigated three parallel physics courses during problem solving sessions at a Chilean University. These courses addressed the same physics curriculum using a different combination of teaching strategies and problems (i.e., well-structured algebra-based and/or ill-structured and open-ended problems). Our research goal is to compare social activity for information seeking and classroom cohesion under this set of teaching and learning conditions (i.e., teaching strategies and problems used), using information seeking as evidence of social interaction and network formation.   

\section{Problem solving and social interactions}
We draw from the literature on physics education research (PER) and the dichotomy between well and ill-structured problems. Generally, problems may be categorized in two different groups, close-ended or well-structured, and open-ended or ill-structured \cite{3_Jonassen}. In physics education, the former normally refers to math-based textbook activities consisting of a well-structured description of a phenomenon, along with the right set of information that allows for unique and known solutions, often times corresponding to magnitudes requested in the problem’s description \cite{4_Heller}. Conversely, ill-structured problems do not show the completeness that textbooks activities do, and students face the challenge of making decisions to develop and create appropriate solutions \cite{5_Fortus}. We believe that under these activities, students are likely to experience different levels of social activity to seek information, motivated by finding accurate understandings or practical responses, as well as novel ideas from their peers.

\section{Social network analysis and tied formation}
Social Network Analysis (SNA) defines a set of theoretical and methodological approaches to understand social systems, its structures, and whether these structures relate to outcomes in a given context. SNA utilizes three basic elements: nodes (e.g., individuals); ties or links between nodes; and graphical representations of networks, which display the pattern of node-to-node relations \cite{6_Putnik}. 
Education research has used SNA to explore for instance, the extent to which social phenomenon influence learning, and whether teaching programs enable community formation. Brewer, Kramer and O'Brien \cite{7_Brewer}, considered contextual features introduced through Modelling Instruction (MI), and explored formation of learning communities as a measure of participation and engagement. Other efforts have tried to provide evidence on the consequences or influences of different social structures over individual or group-level outcomes \cite{2_Stadtfeld,8_Bruun}, where more central or socially embedded individuals tend to enjoy better outcomes. 
A review of network formation in higher education is characterized by student’s capability to find trustworthy peers (i.e., conformity) who show familiar behaviors and attitudes (i.e., homophily) \cite{10_Biancani}, similar features (e.g., gender, place of birth, scores, etc.) \cite{Candia_2018}, or through hierarchical relationships (i.e., distinction of status) \cite{11_McFarland}. Each of these micro-mechanisms for tie formation reflects the extent to which individuals project their attributes into the social space, hoping to find others who reflect  similarities in the case of conformity and homophily, or different ones in order to select weaker or stronger peers that could enable different opportunities. McFarland and colleagues \cite{11_McFarland} suggest that instructional design would facilitate a particular mechanism for network formation, depending on whether students are grouped by cohort, academic level or decide which classes to take. When seeking advice, one would expect individuals to turn to their friends (i.e., similar others), or the ones who they trust the most as a source of valuable information, who may be either someone close in attributes, or a person perceived as valuable for their knowledge, competencies and/or links to others. Furthermore, in this study we explored whether the need for information in solving different types of activities enables community building, and whether status-driven ties govern student’s interactions in a university physics course.

\section{Methods}
Data was collected from a physics course during an academic semester in a University in Northern Chile. A sample of 82 first year students (26 female) from two Engineer majors enrolled in a General Mechanics course. 
Due to the number of enrolled students, three sections were created to address the course curriculum: Class 1 (N = 33, 13 female), Class 2 (N = 23, 8 female) and Class 3 (N = 26, 5 female).  

\subsection{Teaching and learning strategies}
Instructors from Classes 1 and 2 adopted lecture based strategies to address the physic content and solve problems in the classroom. In both settings, instructors responded to students’ questions throughout the sessions by providing direct information on how to understand or solve physics problems. The instructor for Class 3, however, enacted on a more participatory teaching strategy that included frequent social interactions and integration with the student to address emergent and designed problems for the class. Under this strategy, Class 3’s instructor encouraged student-student interactions when they faced conceptual and procedural challenges linked to the course content and problems. During weekly problem solving sessions, Class 1 worked only on well-structured problems, Class 2 used well and ill-structured problems varying every other week, and Class 3 focused on ill-structured problems. 

\subsection{Data collection}
We collected network data the $7^{th}$ week of the academic semester on two variables: \textit{`Information Seeking'} and \textit{`Good Students'} (i.e., whether a given student is perceived as knowledgeable in physics problem solving). We constructed a survey for each class section, which consisted of questions such as: \textit{‘With what frequency did you seek out information from this person to solve physics problems?’} (Never-Always) and, \textit{‘Is this person a good physics student?’} (Yes or No). In addition, we collected Test Scores and Gender for control.

\subsection{Data analysis}
We used the network of Information Seeking to determine cohesion and centrality measures, such as local clustering, betweenness centrality, degree, out-degree and in-degree \cite{12_Borgatti}. Local clustering indicates the probability that a pair of students linked to a focal subject are connected, thus providing a sense of how cohesive is a node’s network. Betweenness centrality is often related to network connectivity, and defined as the count of times a given node (i.e., student) lays in the shortest path between two other nodes. Both metrics enable different interpretations over control of information flow, while local clustering controls for influence over node’s immediate neighbours, betweenness centrality enables control over the information flowing from all pairs of nodes. Degree counts for the number of ties, regardless of direction, whereas out-degree counts for the outgoing ties, and in-degree refers to incoming ties. Finally, to explore social connectivity and interaction of these three classes, we produced five regression models (Table 1) on the three classes, controlling for gender, test scores and good student nomination, in order to predict five network measures: betweenness, local clustering, degree, out-degree and in-degree. Class coefficients come from the differences between the given course and the base category (Class 1). 

\section{Results}
On model for logarithm betweenness centrality, a centrality metric recognized with network connectivity, we observed a positive effect for Class 3 (1.294, p\textless0.05), predicting 23\% of the variance. This result indicates that a combination of ill-structured problems along with encouraging between-group interactions for problem solving enables greater classroom connexion. A positive coefficient between Class 2 and 1 would suggest that the ill-structured problems might affect the social structure of the class. The lack of statistical significance we deemed is due to the small sample size, and therefore, we need more data to reduce the standard error of the estimation. On figure 1A, we observed a clear separation in the probability density function for the logarithm betweenness centrality, suggesting a significant difference between classes. Indeed, when we plotted this variable on student’s test scores (Fig. 1C), we observe the fixed effects for all three courses, with clear distinction between these social systems. 
From model 2 we draw no differences between Class 3 and 1 on local clustering, a measurement of network cohesion. Yet, the test between Class 2 and 1 (-0.169, p\textless0.05) would indicate that close-ended and algebra-based problems yield students to have connected neighbours. Differences on model 3 for logarithm degree centrality between Class 2 and 1 are also significant. Again, the negative regression coefficient shows that the number of social ties measured are higher on the lecture-based Class 1. It is worth mentioning, that even though \textit{good student nomination’s} coefficient is small (0.046, p\textless0.01) on model 3, its significance shows that students will direct their attention towards students perceived as knowledgeable in physics. Non-significant differences between Class 3 and 1 would indicate that both social systems experience similar levels of social activity for information seeking. 
Model 5 supports the interpretation of logarithm degree, but for in-degree centrality. Readers should remember that this measure indicates the number of incoming ties students received from their peers in the classroom, with the purpose of accessing information for solving problems. In other words, this indicates the degree to which students were sought for advice during the problem solving session. Coefficients indicate that in Class 2 students were more connected compared to Class 1 (1.977, p\textless0.01), while this difference is not significant between Class 3 and 1. Again, recognizing skills and knowledge within the social system positively predicts seeking out information, and to whom the information is sought.

\begin{table}[!htbp] \centering 
  \caption{OLS multiple regression models for network centrality measures} 
  \label{table1} 
\tiny 
\begin{tabular}{@{\extracolsep{5pt}}lccccc} 
\\[-1.8ex]\hline 
\hline \\[-1.8ex] 
 & \multicolumn{5}{c}{\textit{Dependent variable:}} \\ 
\cline{2-6} 
\\[-1.8ex] & Log(Betweenness) & Local Clustering & Log(Degree) & Out Degree & In Degree \\ 
\\[-1.8ex] & (1) & (2) & (3) & (4) & (5)\\ 
\hline \\[-1.8ex] 
 Class 2 & 0.761 & $-$0.174$^{**}$ & $-$0.414$^{***}$ & $-$1.399 & $-$1.977$^{***}$ \\ 
  & (0.612) & (0.071) & (0.147) & (1.444) & (0.553) \\ 
  & & & & & \\ 
 Class 3 & 1.294$^{**}$ & $-$0.008 & $-$0.024 & $-$0.774 & $-$0.008 \\ 
  & (0.570) & (0.071) & (0.139) & (1.375) & (0.562) \\ 
  & & & & & \\ 
 Gender  & $-$0.799 & 0.076 & $-$0.103 & $-$1.229 & 0.029 \\ 
  (Female)& (0.480) & (0.062) & (0.124) & (1.222) & (0.486) \\ 
  & & & & & \\ 
 Score & $-$0.158 & 0.017 &  &  & 0.006 \\ 
  & (0.161) & (0.020) &  &  & (0.160) \\ 
  & & & & & \\ 
 Good Student & 0.031 & $-$0.002 & 0.046$^{***}$ & 0.261$^{**}$ & 0.098$^{**}$ \\ 
 Nominations& (0.044) & (0.006) & (0.011) & (0.112) & (0.044) \\ 
  & & & & & \\ 
 Constant & 1.804$^{**}$ & 0.374$^{***}$ & 1.614$^{***}$ & 2.424 & 3.658$^{***}$ \\ 
  & (0.869) & (0.114) & (0.158) & (1.560) & (0.896) \\ 
  & & & & & \\ 
\hline \\[-1.8ex] 
Observations & 51 & 77 & 82 & 83 & 79 \\ 
R$^{2}$ & 0.234 & 0.112 & 0.358 & 0.117 & 0.309 \\ 
Adjusted R$^{2}$ & 0.149 & 0.049 & 0.324 & 0.071 & 0.261 \\ 

\hline 
\hline \\[-1.8ex] 
\textit{Note:}  & \multicolumn{5}{r}{$^{*}$p$<$0.1; $^{**}$p$<$0.05; $^{***}$p$<$0.01} \\ 
 & \multicolumn{5}{r}{Base categories: Class 1 and Gender (Male)} \\ 
\end{tabular} 
\end{table} 

\begin{figure*}[!t]
 \centering 
  \includegraphics[width=1\textwidth]{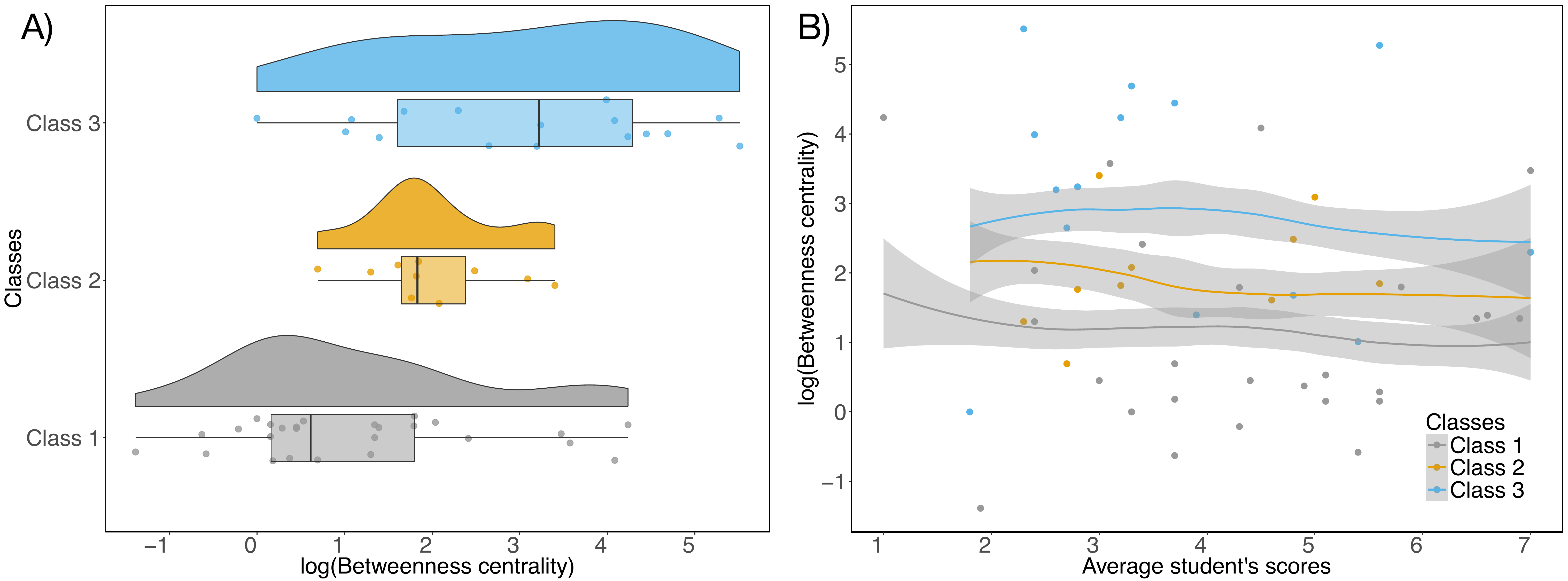}
\caption{A) Density-plot, box-plot and scatter-plot for the logarithm of betweenness centrality of each analyzed class. B) Logarithm of betweenness centrality v/s average student's score, the lines represent the linear regressions for each class. Dots represent the values for individual students.}
 \label{figura1}
\end{figure*}

\section{Discussion and conclusions}
It may not be surprising that the teaching strategy enacted by the instructor in Class 3 yields higher classroom cohesion (Model 1). This result is a promising support for the design and use of ill-structured physics problems, along with techniques for peer-peer support as appropriate techniques for community building. However, the lack of significance in local clustering between Class 3 and 1 suggest that both classes enable similar opportunities for local connectivity and triadic ties. This would imply that on Class 1 and 3, students were capable of seeking out information on peers who are connected to each other, thus facilitating information flow within clusters. In contrast, higher betweennes centrality on Class 3 indicates overall higher connectivity, and more control over the flow of information available to all students in the system.  
Moreover, and because differences were obtained in favour of Class 1 over 2 on degree centrality, out-degree and in-degree, we argue that having unique solutions (i.e., well-structured problems) may motivate students to pursue information regarding the correctness of their procedures and outcomes, a social phenomenon that students may not engage in with ill-structured problems, given the uniqueness of their solutions. This interpretation is appropriate given that instructors in Class 1 and 2 behaved similarly in guiding problem solving sessions, by providing direct information to students’ inquiries. As previously mentioned, Class 2 used ill-structured problems every other week, activities that require idea generation, thus allowing multiple possible solutions. Consequently, solving activities with multiple solutions may dampen the need of seeking out information, as others’ solutions may not necessarily relate to the solution path decided by a given group of students. 
These results are promising for finding novel teaching and learning techniques for university-level science courses. However, further research needs to be conducted to explore the ways in which student groups tackle well and ill-structured problems, and to reveal possible differences in the way these activities enable community development, and under which mechanisms they do so.

\subsection*{Acknowledgments}
This material is based upon work supported by the AAPT E. Leonard Jossem International Education Fund. Any opinions, findings, and conclusions or recommendations expressed in this material are those of the author(s) and do not necessarily reflect the views of the American Association of Physics Teachers.

\end{document}